\documentclass[twocolumn]{aastex62}
\usepackage{graphicx}% Include figure files
\usepackage{dcolumn}% Align table columns on decimal point
\usepackage{bm}% bold math
\usepackage{epstopdf,color,verbatim}
\usepackage{hyperref}
\usepackage{xspace}
\usepackage{amsmath}
 
\graphicspath{{figures/}{./}}

\newcommand{\fermi}{\textit{Fermi}\xspace}
\newcommand{\eqb}{\begin{eqnarray}}
\newcommand{\eqe}{\end{eqnarray}}

\definecolor{royalblue}{RGB}{65,105,225}

\graphicspath{{./}{figures/}}

\received{}
\revised{}
\accepted{\today}
\submitjournal{ApJ}

\shorttitle{Plasmoid flares seen with \fermi}
\shortauthors{Meyer et al.}
\defcitealias{Christie2019}{CPS19}
\defcitealias{Meyer2019}{MSB19}

\begin{document}

\title{The observability of plasmoid-powered $\gamma$-ray flares with the \fermi Large Area Telescope}

\correspondingauthor{Manuel M.~Meyer}
\email{manuel.e.meyer@fau.de}

\correspondingauthor{Maria Petropoulou}
\email{mpetropo@phys.uoa.gr}

\author[0000-0002-0738-7581]{Manuel M.~Meyer}
\affil{Erlangen Centre for Astroparticle Physics, University of Erlangen-Nuremberg, Erwin-Rommel-Str. 1, D-91058 Erlangen, Germany
}
 
\author[0000-0001-6640-0179]{Maria Petropoulou}
\affil{Department of Physics, National and Kapodistrian University of Athens, Panepistimiopolis, GR 15783 Zografos, Greece}
\affil{Department of Astrophysical Sciences, Princeton University \\
4 Ivy Lane, Princeton, NJ 08544, USA}

\author{Ian M.~Christie}
\affiliation{Center for Interdisciplinary Exploration \& Research in Astrophysics (CIERA), Northwestern University \\
Evanston, IL 60208, USA}



\begin{abstract}
The exact mechanism for the production of fast $\gamma$-ray variability in blazars remains debated. Magnetic reconnection, in which plasmoids filled with relativistic particles and magnetic fields are formed, is a viable candidate to explain the broadband electromagnetic spectrum and variability of these objects. 
Using state-of-the-art magnetic reconnection simulations, we generate realistic $\gamma$-ray light curves that would be observed with the \textit{Fermi} Large Area Telescope. 
A comparison with observed $\gamma$-ray flares from flat spectrum radio quasars (FSRQs) reveals that magnetic reconnection events lead to comparable flux levels and variability patterns, in particular, when the reconnection layer is slightly misaligned with the line of sight. 
Emission from fast plasmoids moving close to the line of sight could explain the fast variability on the time scales of minutes for which evidence has been found in observations of FSRQs.
Our results motivate improvements in existing radiative transfer simulations as well as dedicated searches for fast variability as evidence for magnetic reconnection events. 
\end{abstract}



\section{Introduction} \label{sec:intro}
Blazars are a subclass of active galactic nuclei (AGNs) with relativistic plasma outflows (jets) powered by accretion onto their supermassive black hole and closely aligned to our line of sight. These AGNs are the most powerful persistent astrophysical sources of nonthermal electromagnetic radiation in the universe, with bolometric luminosities of $\sim10^{43}-10^{48}$~erg s$^{-1}$ \citep[e.g.,][]{Ackermann2015}. The spectral energy distribution (SED) of blazar emission is broadband, spanning about 
15 decades in energy
from radio frequencies up to high-energy $\gamma$-rays  \citep[for a recent review, see][]{Dermer2016}. 

Blazar variability, which refers to flux fluctuations away from an average value, is frequency dependent and manifests in a variety of timescales that can be as long as the time span of the data under study and as short as the detector's signal-to-noise ratio allows. 
Already from the era of the Energetic Gamma Ray Experiment Telescope, it has been known that $\gamma$-rays may vary on day-long timescales \citep[see,][for 3C 279]{Kniffen1993}. Since then, flux doubling timescales as short as a few minutes have been detected in several blazar flares\footnote{Although there is no unique definition, broadly speaking, flares constitute changes in flux within a factor of a few that are limited in time.} with Cerenkov telescopes  at very high energies ($>100$~GeV) \citep[see, e.g.,][]{Aharonian2007, Albert2007, Arlen2013, MAGIC2020_Mrk421} and in the 0.1--300 GeV energy range with the \fermi Large Area Telescope %-LAT 
\citep[LAT;][]{Aleksic2011, Ackermann2016, Shukla2018}. The brightest $\gamma$-ray flares of six flat spectrum radio quasars\footnote{These are traditionally differentiated from other blazar subclasses by the equivalent width of lines in their optical  spectra \citep{Urry1995}, but see also \citet{Padovani2017} for a recent review.} (FSRQs) within $\sim10$~yr of \fermi-LAT observations have been recently studied by \citet[][hereafter \citetalias{Meyer2019}]{Meyer2019}. 
Evidence for sub-hour timescales (as short as minutes) has been presented for two FSRQs in the sample (3C 279 and CTA 102) at a $\sim2\,\sigma$ post-trial significance.
For two additional sources (PKS 1510-089 and 3C 454.3) decay times on timescales similar or less than the satellite's orbit (95 minutes) were found \citepalias[see Figure~6 in][]{Meyer2019}.

These short-duration $\gamma$-ray flares are of particular importance as they are a manifestation of the underlying physical processes that power the multi-timescale temporal variability in blazars and can constrain models of $\gamma$-ray production in jets
\citep[e.g.,][]{Giannios2010, Nalewajko2011, Barkov2012, Narayan2012, Aleksic2014, Ackermann2016, Aharonian2017, Petropoulou2017}. The observed duration of several minutes is too short as compared to the natural scale of the blazar engine, namely the light-crossing time of the central black hole, $t_{\rm g}=2 GM_{\rm BH}/c^3 \simeq 10^4\left(M_{\rm BH}/10^9 M_{\odot}\right)\,\mathrm{s} \simeq 160\left(M_{\rm BH}/10^9 M_{\odot}\right)$~minutes. Thus, the emitting region of the observed minute-long flares should either be of sub-horizon scale \citep[e.g.,][]{Aleksic2014} or should move relativistically in the jet \citep{Begelman2008, Giannios2009,Narayan2012}. For ultra-rapid teraelectronvolt flares, in particular, $\gamma\gamma$ opacity places a lower limit\footnote{This is valid as long as the low-energy synchrotron emission also originates from the same region as the teraelectronvolt $\gamma$-rays.} on the Doppler factor of the emitting region, e.g., $\delta \gtrsim 50$ \citep[e.g.,][]{Begelman2008,Finke2008}.   
These Doppler factors are typically much larger than those inferred from GHz observations of parsec-scale blazar jets \citep[e.g.,][]{Liodakis2018, Finke2019}. However, this requirement on the jet's Lorentz factor $\Gamma_j$ can be relaxed if the emitting region moves relativistically within the jet with Lorentz factors $\Gamma_{\rm co}$ of a few. With this, the resulting Doppler factor is then $\delta \sim \Gamma_{\rm co} \Gamma_j$, which can easily exceed $40$  for typical values of $\Gamma_j$.

Magnetic reconnection -- a process where magnetic energy is transferred to the plasma bulk motion and acceleration of particles -- offers a natural way of producing compact magnetized structures, the so-called plasmoids, moving at relativistic speeds with respect to the co-moving frame of the jet \citep[e.g.,][]{Loureiro2007, Uzdensky2010, Loureiro2012}. Plasmoids, which contain nonthermal particles, can accelerate due to magnetic stresses up to a terminal Lorentz factor $\Gamma_{\rm co}\approx 3\left(\sigma/10\right)^{1/2}$, where $\sigma\gtrsim1$ is the jet's plasma magnetization \citep{Lyubarsky2005, Sironi2016}. In addition to bulk acceleration, plasmoids grow in size by accreting plasma either through the secondary current sheets between plasmoids or through mergers with other neighboring plasmoids. Because of their properties, plasmoids have been suggested as candidate sites for the production of short-duration $\gamma$-ray flares from blazars \citep{Giannios2013, Petropoulou2016}.

A single reconnection event can lead to the production of a plasmoid chain, namely a series of plasmoids with different sizes and bulk speeds \citep{Shibata2001, Uzdensky2010,Huang2012, Loureiro2012, Sironi2016}. \citet{Petropoulou2018} have investigated the various physical processes (e.g., growth, acceleration, and mergers) that shape the size and momentum
distributions of the plasmoid chain using a Monte Carlo approach. The size distribution of plasmoids was found to be a power law extending up to a size of $\sim0.2$ 
of the reconnection layer's half length $L$ (measured in the jet rest frame), with a slope that is only weakly dependent on the magnetization.
Given that each plasmoid can power a $\gamma$-ray flare with luminosity and duration related to its size and bulk Lorentz factor \citep{Giannios2013, Petropoulou2016}, the plasmoid chain should produce numerous flares with a wide range of luminosities and durations \citep[see, e.g., Figure 17 in][]{Petropoulou2018}.

By coupling recent two-dimensional (2D) particle-in-cell simulations of relativistic reconnection with a time-dependent radiative transfer code,  
\citet[][hereafter \citetalias{Christie2019}]{Christie2019} computed the nonthermal emission from a chain of plasmoids formed during a single reconnection event. The authors showed that, at any given time, an observer receives radiation from a large number of plasmoids, which are present in the layer, with different sizes and Doppler factors.  Plasmoids that move with mildly relativistic speeds (in the jet frame) and have intermediate sizes (e.g., $\gtrsim 0.05\, L$) were found to contribute significantly to the overall emission \citepalias[see Figure 9 in][]{Christie2019}. The superposition of their emission can result in a slow varying component of the light curve, i.e., an envelope of emission, as first proposed by \citet{Giannios2013}, while smaller plasmoids (with sizes $\lesssim 0.01\,L$), which typically move with relativistic speeds (i.e., $\Gamma_{\rm em} \approx \sqrt{\sigma}$), can result in luminous and ultra-rapid flares that show up on top of the envelope emission \citepalias[see Figure 9 in][]{Christie2019}.

The goal of this paper is to investigate if various features of the theoretical light curves, such as the ultra-rapid $\gamma$-ray flares predicted by the reconnection model, are detectable with the typical \fermi-LAT observations. The LAT is a pair-conversion telescope that detects $\gamma$-rays in the energy range between 30\,MeV and beyond 300\,GeV and surveys the full sky every $\sim3$~hr \citep{2009ApJ...697.1071A}.

In our analysis, we select two prototypical FSRQs, namely 3C~273 and 3C~279.
Both sources are bright $\gamma$-ray emitters that have shown strong outbursts at $\gamma$-ray energies in the past \citep[e.g.,][]{Rani2013, Ackermann2016}. The different redshifts of the sources ($z=0.158$ and $z= 0.536$ for 3C~273 and 3C~279, respectively) and the different strengths of external photon fields (i.e., disk and broad-line region, BLR) enable us to study their impact on  the theoretical light curves.

This paper is structured as follows. In Section~\ref{sec:methods} we present our methodology for creating artificial \fermi light curves based on the light curves computed from the reconnection model. In Section~\ref{sec:results} we present our results on the artificial \fermi light curves and resulting power spectra. We also compare these results with predictions of magnetic reconnection simulations for BL Lac objects.
% We also note on predictions of the magnetic reconnection simulations for BL Lac objects. 
In Section~\ref{sec:models} we review  alternative models for fast $\gamma$-ray variability, and make a qualitative comparison with the magnetic reconnection model. We continue in Section~\ref{sec:discussion} with a discussion of our results, and conclude in Section~\ref{sec:conclusion}. Throughout this paper, we adopted a cosmology with $\Omega_M=0.3$, $\Omega_\Lambda=0.7$,  and $H_0=70$~km~s$^{-1}$ Mpc$^{-1}$. 

\section{Methods}\label{sec:methods}
First, we briefly discuss the parameters of the theoretical light curves from the reconnection model. We then continue with a detailed description of the methods used to create  artificial \fermi-LAT light curves based on the reconnection model. 

\begin{deluxetable*}{cccc  cccc}
% \colnumbers
\centering
\tablecaption{Parameters used for the computation of theoretical $\gamma$-Ray light curves from magnetic reconnection in blazar jets. \label{tab:models}
}
\tablewidth{0pt}
\tablehead{
\colhead{Model}  & \colhead{$\Gamma_j$} &    \colhead{$\theta_{\rm obs}$ (deg)} & \colhead{$\theta^\prime$ (deg)} & \colhead{$z_{\rm diss}$\tablenotemark{a} (cm)} & \colhead{$L_{\rm ext}$\tablenotemark{b}  (erg s$^{-1}$)} & \colhead{$L_j$\tablenotemark{c} (erg s$^{-1}$)}
& \colhead{Blazar}
  }
\startdata 
A & 12   &   0 & 0 &  $6\times 10^{17}$ & $4\times10^{45}$ & $10^{47}$ & 3C~273 \\
B & 24  &  0.2 & 0 &  $1.2\times10^{18}$  &$4\times10^{45}$ & $5\times 10^{47}$ & 3C~273  \\
C & 24  &   0 & 30 & $1.2\times10^{18}$ & $4\times10^{45}$ & $5\times 10^{47}$ & 3C~273 \\
D & 24  &   0 & 0 & $1.2\times10^{18}$  & $10^{46}$ & $5\times 10^{47}$ & 3C~279  \\
\enddata
\tablenotetext{a}{Dissipation distance, estimated as $z_{\rm diss}\approx \Gamma_j L$, for a conical jet with half-opening angle $\theta_j \approx \Gamma_j^{-1}$. }
\tablenotetext{b}{Lower limit on the bolometric luminosity of the external radiation field.}
\tablenotetext{c}{Absolute power of a two-sided jet \citepalias[see Equation 1 in][]{Christie2019}.}
\tablecomments{Other parameters used are the half length of the reconnection layer $L=5\times10^{16}$~cm, plasma magnetization $\sigma=10$, magnetic field strength of unreconnected (upstream) plasma $B=5$~G at $z_{\rm diss}$, plasmoid-averaged co-moving energy density of nonthermal pairs $u^\prime_e= 2.2$ erg~cm$^{-3}$ \citepalias[see Equation 6 in][]{Christie2019}, minimum (maximum) particle Lorentz factor (averaged over all plasmoids) $\gamma_{\min} = 94$ ($\gamma_{\max}= 5 \times 10^3$), and slope of injection power-law particle spectrum $p=2.1$.
}
\end{deluxetable*} 

\begin{figure*}
    \centering
    \includegraphics[width = .99\linewidth]{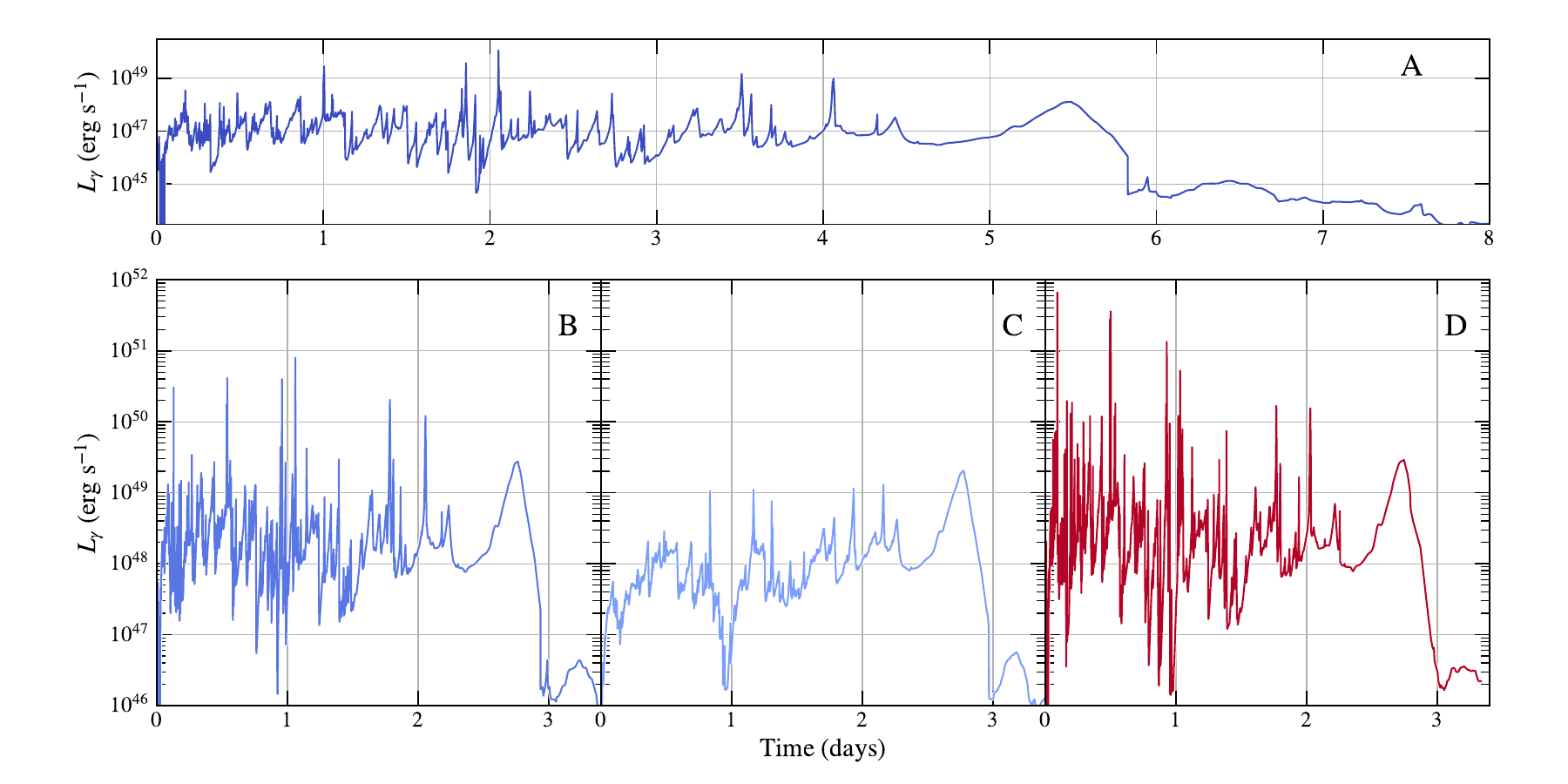}
    \caption{Theoretical $\gamma$-ray  light curves in the $0.1 - 300$~GeV energy range produced by hundreds of plasmoids formed in a reconnection layer in the jet. The light curves are computed using the numerical code of \citetalias{Christie2019}
    for the model parameters considered in Table~\ref{tab:models}. Variations in the duration and variability of the entire reconnection event are the result of changing the bulk jet Lorentz factor and its orientation with respect to both the jet axis and the observer.}
    \label{fig:theo-lc}
\end{figure*} 

\subsection{Theoretical light curves}\label{sec:theory}
\citetalias{Christie2019} computed the multiwavelength spectra and light curves powered by a chain of plasmoids formed in magnetic reconnection for a generic blazar belonging to the BL Lac or FSRQ subclasses. Within the magnetic reconnection scenario for blazar emission, \citetalias{Christie2019} showed that the transition from a BL Lac to an FSRQ type of blazar can be achieved by (i) decreasing the magnetization in the jet's dissipation region from $\sigma>10$ to $\sigma \sim3$, (ii) increasing the number of electrons per proton in the jet from one to a few, (iii) increasing the jet power, and (iv) increasing the strength of the external radiation fields.

In this paper, we investigate the $\gamma$-ray timing properties of the magnetic reconnection model, as seen by \fermi-LAT, using two prototypical FSRQs as test beds. For this purpose, 
we compute theoretical light curves using the numerical code of \citetalias{Christie2019}.
We adopt the ``vanilla'' FSRQ model of \citetalias{Christie2019}, listed as Model A in Table~\ref{tab:models}. The variability properties of the escaping radiation from the reconnection layer are sensitive to changes in the Doppler factor, $\delta_p$, of individual plasmoids formed in the layer \citepalias[see e.g., Figs.~7 and 8 in][]{Christie2019}. The Doppler factor depends on the bulk Lorentz factor $\Gamma_j$, the observer's angle $\theta_{\rm obs}$, and the angle between the layer and the jet axis as measured in the jet frame $\theta^\prime$ (for detailed explanation of the model parameters, we refer the reader to \citetalias{Christie2019}). We therefore add to our analysis three additional models (Models B-D) that differ in the above-mentioned parameters. All models are generic, i.e., not meant to fit the multiwavelength emission of the chosen sources during specific epochs. 

In all models the external radiation field is considered to be isotropic in the AGN rest frame. It is modeled as a graybody with effective temperature $T=10^4$~K and total energy density $u_{\rm ext}=0.037$~erg cm$^{-3}$ (assuming a covering fraction $0.1$), occupying a spherical region of radius $R_{\rm ext}$
\citep{Ghisellini1996}.
Magnetic reconnection is assumed to take place within this radius, so that the energy density in the co-moving frame of plasmoids appears boosted as $\approx \Gamma_p^2 u_{\rm ext}$, where $\Gamma_p = \Gamma_j \Gamma_{\rm co} (1 +\beta_j \beta_{\rm co} \cos\theta')$ is the plasmoid Lorentz factor in the galaxy's frame, $\Gamma_{\rm co}$ is the plasmoid Lorentz factor in the jet frame, and $\beta_{j \rm (co)}=\sqrt{1+\Gamma_{j (\rm co)}^{-2}}$.  For  a conical jet with half-opening angle $\theta_j \approx \Gamma_j^{-1}$, the dissipation distance can be estimated as $z_{\rm diss}\approx \Gamma_j L$. In all models we make the implicit assumption that $z_{\rm diss} \lesssim R_{\rm ext}$, which translates into a lower limit on the bolometric luminosity of the external radiation field $L_{\rm ext}\approx (48\pi/17) R_{\rm ext}^2 c u_{\rm ext}$. In Models A-C, $L_{\rm ext}\gtrsim 4\times10^{45}$~erg s$^{-1}$, which is consistent with the BLR luminosity of 3C~273 \citep{Peterson2004,Vasudevan2009}.  In Model D, $L_{\rm ext}\gtrsim 10^{46}$~erg s$^{-1}$, while the luminosity of the thermal emission component in 3C~279 is 
$\sim 3\times10^{45}$~erg s$^{-1}$, which translates to  
$L_{\rm BLR}\sim3\times10^{44}$~erg s$^{-1}$ \citep{Celotti1997, Pian1999}\footnote{
We have converted the luminosities provided by \citet{Celotti1997} and \citet{Pian1999} using updated cosmological parameters (see the end of Section~\ref{sec:intro}) and redshift for 3C~279.
}. Thus, the properties of the external radiation field used in our model for 3C 279 are not consistent with the average properties of its BLR emission. We discuss the implications of our model for 3C 279 in Section~\ref{sec:discussion}. 

The theoretical light curves used as our baseline models for computing artificial \fermi light curves are displayed  in Figure~\ref{fig:theo-lc}. The differences in duration and luminosity between models shown in Figure~\ref{fig:theo-lc} arise purely from changes in $\Gamma_j$, $\theta_{\rm obs}$, and $\theta^\prime$. For perfect alignment of the layer with the observer  ($\theta_{\rm obs} = \theta^\prime = 0^\circ$), the duration of a reconnection event is $\propto L/ \Gamma_j$ while  the peak luminosity of a $\gamma$-ray flare powered by a specific plasmoid is $\propto \Gamma_j^{2 + p} \, L^{2}$, where $p$ is the power-law slope of the particle injection spectrum (see the Appendix for details). Models B-D are on average more luminous than Model A  because $\Gamma_j$ is two times larger.
% due to the twice as high value of $\Gamma_j$. 
For the same reason, the total duration of the light curve in Models B-D is similar. However, the variability on short timescales between Models C and B (or D) is still affected by the different angles used, because these have an impact on the Doppler factor of the faster plasmoids in the layer (e.g., compare the first 1-hr segment of the light curves in Models B and C).

\subsection{Production of artificial \fermi light curves}
\label{sec:fakelc}

Our goal is to produce the most realistic source light curves possible
in order to determine whether and which characteristics of the theoretical light curves are retained in an actual \fermi-LAT observation. 
To be as close as possible to an actual observation, we have developed the following procedure to simulate and analyze observations:
\begin{enumerate}
    \item Select an actual FSRQ and some time window.     
    \item Conduct a standard \fermi analysis of optimizing source spectral
    parameters in a given region of interest (ROI) and derive a light curve with a specific temporal binning.
    \item Replace the central source with the prediction of the theoretical model in each time bin, add a baseline flux, and simulate the modified ROI\footnote{An ROI is usually simulated by drawing a random Poisson realization from the sky map containing the expected number of counts from each source. Instead, we use the number of expected counts directly as our simulation. 
    This so-called Asimov data set yields results close to the median of many random Monte Carlo simulations \citep{2011EPJC...71.1554C}.}.
    \item Conduct a standard \fermi analysis on the simulated ROI and reconstruct the source light curve. \end{enumerate}

For each of the two considered blazars 3C~273 and 3C~279, we choose time windows that encompass reported $\gamma$-ray flares, taken from \citetalias{Meyer2019}. These are listed in Table~\ref{tab:timewindows}. 

\begin{deluxetable}{l cc}
% \colnumbers
\centering
\tablecaption{Time windows used for the generation of artificial \fermi light curves for the  selected sources.
\label{tab:timewindows}
}
\tablewidth{0pt}
\tablehead{
\colhead{Source}  & \colhead{Start time $t_0$ (MJD)} &  \colhead{End time (MJD)} 
  }
\startdata 
3C~273 & 55,094.69 & 55,104.69 \\
3C~279 & 58,140.66 & 58,150.66 \\
3C~279 & 57,189.08 & 57,189.11
\enddata
\tablecomments{
The second time window for 3C~279 encompasses the orbit during the brightest period of the $\gamma$-ray flare observed in June 2015.
This will be used in Section~\ref{sec:results} to compare to the highest luminosities predicted by the magnetic reconnection simulations on short time scales.
}
\end{deluxetable}
For the \emph{Fermi} analysis (step 2 above), we analyze events of the \texttt{P8V3 Source} class with energies between 100\,MeV and 316\,GeV that have arrived at a zenith angle $\leqslant 90^\circ$, in order to minimize contamination with Earth limb photons. 
For the analysis, we use \textsc{fermipy}\footnote{\url{https://fermipy.readthedocs.io}} version 0.19.0 \citep{Wood2017} and run the standard \textsc{fermitools}\footnote{\url{https://fermi.gsfc.nasa.gov/ssc/data/analysis/}}, version 1.2.23.
Gamma~rays within a $15^\circ \times 15^\circ$ ROI centered on the given FSRQ are considered and the standard quality cuts of \texttt{DATA\_QUAL$>$0} and \texttt{LAT\_CONFIG==1} are applied.
Periods of bright $\gamma$-ray bursts and solar flares that have been detected with a test statistic (TS) $> 100$ are excluded\footnote{The $\mathrm{TS}$ value is defined as twice the difference between the log-likelihoods for the model with and without the source included \citep{1996ApJ...461..396M}.}.
In our ROI model, we include all sources listed in the fourth \emph{Fermi} point source catalog \citep[4FGL;][]{4FGL} that have a separation less than $30^\circ$ from the ROI center.
Additionally, we include templates for the isotropic diffuse and Galactic diffuse emission\footnote{\url{https://fermi.gsfc.nasa.gov/ssc/data/access/lat/BackgroundModels.html}}.

We first derive a best-fit model for the ROI over 10\,yr of \fermi-LAT observations between 2008 August 4 and 2018 August 4. 
For this 10~yr period, all spectral parameters are free to vary for sources within $5^\circ$ from the ROI
center, and for sources between $5^\circ$ and $10^\circ$, only the normalization is left as a free parameter.
For sources with an angular distance larger than $10^\circ$, all spectral parameters are fixed to their 4FGL values. Additionally, the parameters of sources 
%whose predicted number of photons is less than $10^{-3}$ or 
whose detection significance is $\mathrm{TS} \leqslant 1$ are frozen.  
The normalizations of the background templates are additional free parameters.
We also test if additional sources are present in the ROI by computing a $\mathrm{TS}$ map. We iteratively add additional sources for which $\sqrt{\mathrm{TS}} \geqslant 5$.

After the optimization of the 10 yr period, 
we derive average ROI models for the considered time ranges leaving the same spectral parameters free to vary as before. 
Using these ROI models as a baseline, we derive light curves for the two FSRQs with 3~hr temporal binning. 
In each time bin, we re-optimize the spectral parameters of the central FSRQ and leave the other source parameters are fixed.
Time bins with less than three observed counts are skipped since the parameter optimization usually fails for such a low number of observed photons.

Next, we remove the central real FSRQ from the source model and replace it with an `` artificial'' one based on the prediction of the magnetic reconnection model (step 3 above). 
We average the theoretical light curve to obtain the same temporal binning as the observed light curve and convert the luminosity to flux using the source's luminosity distance and assuming, for the spectrum, a power law with a photon index $\Gamma = 2$ between 100\,MeV and 316\,GeV. 
The chosen spectrum is consistent with the average spectral properties of the vanilla FSRQ model from \citetalias{Christie2019}. 
For each source the same pivot energy as in the 4FGL is used. 
The flux prediction of the theoretical model  refers to a flaring episode caused by a single reconnection event in the blazar jet; it is not designed to explain the long-term quasi-steady state $\gamma$-ray blazar emission, 
here defined as the quiescent background (or QB for short). Thus, we add a quiescent flux to the source model, which, for simplicity is assumed to have the same spectral shape. 
The QB value is taken from \citetalias{Meyer2019}, who found a QB level for the full 9.5~yr long light curve of $3.2\times10^{-7}\,\mathrm{cm}^{-2}\,\mathrm{s}^{-1}$ for 3C~279 and $1.9\times10^{-7}\,\mathrm{cm}^{-2}\,\mathrm{s}^{-1}$ for 3C~273, respectively. 
The spectrum is additionally multiplied with the exponential absorption on the extragalactic background light (EBL) using the model of \citet{2011MNRAS.410.2556D} and on the external photon field. 
In order to use an explicit model for the external photon field, we assume that it is given by the BLR model of \citet{Finke2016}.
\citet{Finke2016} modeled the BLR as a collection of infinitesimally thin rings (or shells) and each ring emits a monochromatic line with a fixed relative luminosity.  
By setting the H$\beta$ luminosity in the BLR model to $1.3\times10^{44}\,\mathrm{erg}\,\mathrm{s}^{-1}$ ($3.3\times 10^{44}\,\mathrm{erg}\,\mathrm{s}^{-1}$) for Models A-C (Model D)
we reproduce the external photon field luminosity used in the magnetic reconnection simulations (Table~\ref{tab:models}). 
The $\gamma$-ray emitting region is placed at a distance $z_\mathrm{diss}$ as reported in Table~\ref{tab:models}. For the black hole masses of 3C 273 and 3C 279, we adopt, respectively, the values $10^{8.92}M_{\odot}$ and $10^{8.23} M_{\odot}$ from  \citet{2006ApJ...637..669L}.
% the same values as in \citetalias{Meyer2019}.

Finally, we repeat the light-curve analysis in the same way as was done with the observed data (step 4) but now with the new flaring central source from the reconnection model.

\section{Results} \label{sec:results}
In this section, we compare the observed light curves with those obtained from the simulations as well as the corresponding power spectral densities (PSDs). 
We do not attempt to reproduce the observed light curves one to one but rather focus on the overall structure, variability time scales, and the PSDs.  To our knowledge, this is the first time that such a comparison has been performed. 
 
\begin{figure*}[tbh]
    \centering
    \includegraphics[width=.99\linewidth]{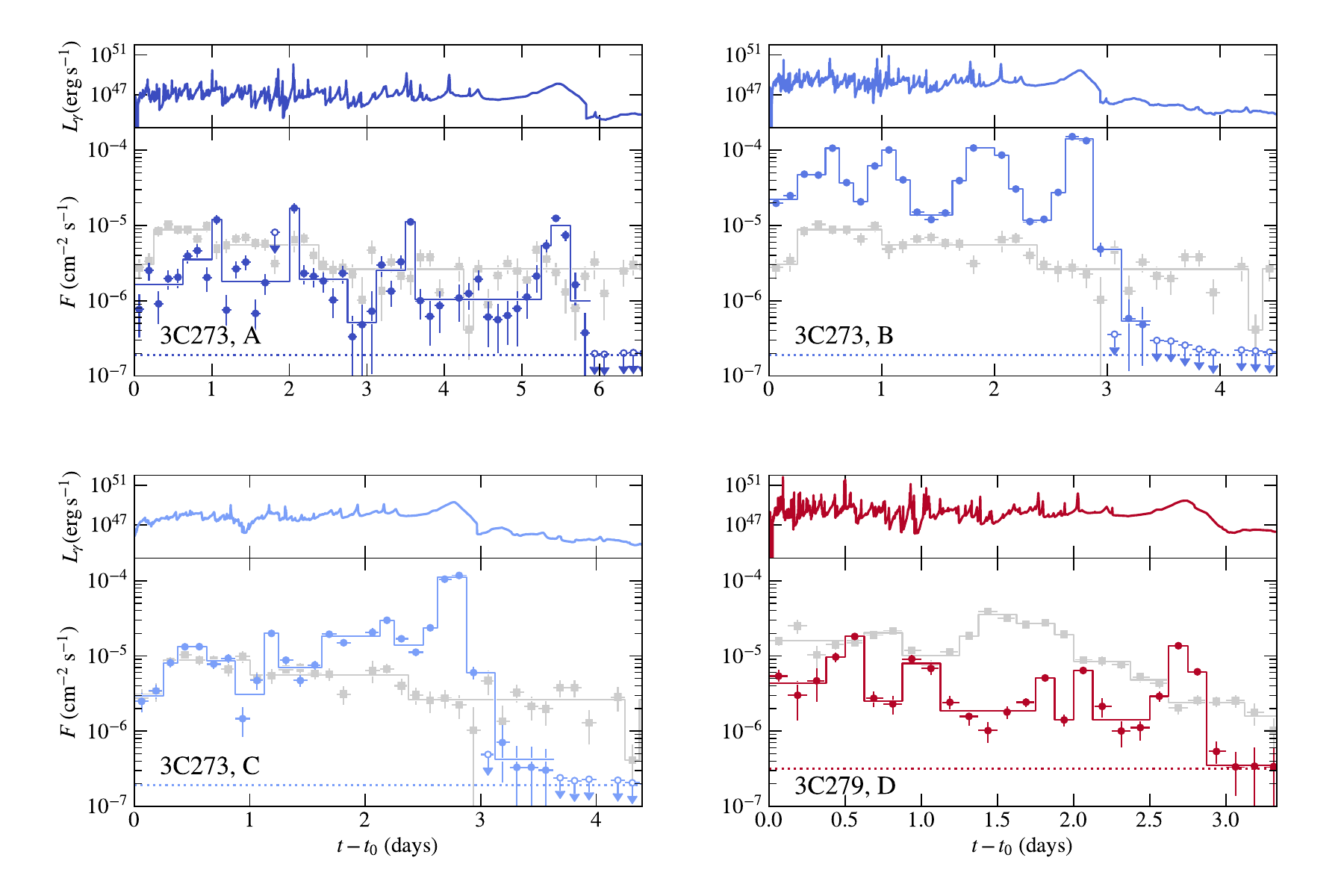}
\caption{
    Theoretical, simulated, and observed  $\gamma$-ray light curves in the 0.1-300\,GeV energy band for each theoretical model discussed in  Section~\ref{sec:theory}. 
    The theoretical light curves are shown in the top panels of each figure, while the observed and simulated  light curves with 3~hr binning are shown in the bottom panels.
    The blue (red) points show the simulated light curve as would be observed with \fermi~LAT from 3C~273 (3C~279) in the given observation windows. 
    The dotted lines show the assumed QB flux level. 
    The gray points show the actually observed light curves. 
    Solid lines show the Bayesian block representation of the observed and simulated light curves. 
    Missing data points in the observed light curves are due to low counts in that time bin, which are in turn due to either low activity or low exposure.
    The light curves are plotted with respect to the start time of the observation, $t_0$, see Table~\ref{tab:timewindows}.
    \label{fig:lcs}}
\end{figure*}
\subsection{Light curves}
The theoretical and artificial \fermi light curves computed for the four models discussed in Section~\ref{sec:theory} are displayed in Figure~\ref{fig:lcs}. The artificial light curves are compared against those of two \fermi-detected $\gamma$-ray flares from 3C~273 (Models A-C) and 3C~279 (Model D).

The flux of the artificial light curves in Models A and C
is broadly consistent with the observed flux of the flare under study, whereas Model B predicts $\sim10$ times higher flux. 
The flux prediction of Model D is not sufficient to reproduce the brightest flare observed from 3C~279.
However, the model-predicted fastest and brightest flares (see spikes on top panel) can produce similar levels as the observed flux and we return to this point later in this section.

As indicated by the Bayesian block representation \citep{2013ApJ...764..167S} of the artificial light curves, much of the short-time variability that is visible in the theoretical light curves is lost, due to the chosen binning (here, 3~hr) and averaging. In fact, none of the sharp spikes of the theoretical light curves is individually recovered in the reconstructed light curves, because they all have durations less than the adopted time bin of 3~hr. As a result, most of the features seen in the reconstructed light curves as individual flares (see, e.g., the first and second blocks in the Bayesian block representation of the light curve for Model  A) are in reality the superposition of multiple unresolved flares powered by individual plasmoids in the layer. Only the longest duration flare, which is produced by the single largest plasmoid in the layer, is recovered in the reconstructed light curves of all models  (see, e.g., the last Bayesian block of reconstructed light curves in Models B and C).

We next compare the minimum variability time scale of the simulated and the observed light curves, which is defined as \citep{1999ApJ...527..719Z}
\begin{equation}
    \mathrm{min}(t_\mathrm{var}) = \underset{i,j}{\mathrm{min}}\left(\frac{F_i + F_j}{2}\left|\frac{t_i - t_j}{F_i - F_j}\right|\right),
\end{equation}
where $F_i$, $F_j$ are the fluxes at times $t_i$, $t_j$ for all pairs of flux points $i, j$.
The values for the models and the observed light curves are reported in Table~\ref{tab:tvar}.
It can be seen that the predicted minimum variability time scales are broadly consistent with the observed ones, albeit shorter, especially for Models B and C.
These models yield light curves with more structure than observed in the specific time window of 3C~273. 

\begin{deluxetable}{cc cc }
% \colnumbers
\centering
\tablecaption{Minimum variability timescales computed for the artificial light curves and for the actually observed LAT light curves of 3C~273 and 3C~279.
\label{tab:tvar}
}
\tablewidth{0pt}
\tablehead{
\colhead{Model} & 
\colhead{Source} & 
\colhead{$\min(t_\mathrm{var})$ (minutes)} &
\colhead{$\min(t_\mathrm{var})$ (minutes)} \\ 
& & \colhead{Artificial} & \colhead{Obs.} 
}
\startdata 
A & 3C~273 & $102.1 \pm 16.5$ & $140.4 \pm 75.8$  \\
B & 3C~273 & $96.7 \pm 3.3 $ & $140.4 \pm 75.8$ \\
C & 3C~273 & $99.7 \pm 3.7$ & $140.4 \pm 75.8$ \\
D & 3C~279 & $107.2 \pm 10.8$ & $215.9 \pm 91.4$ \\
\enddata
\end{deluxetable}

\begin{figure*}[tbh]
    \centering
    \includegraphics[width = 0.99\linewidth] {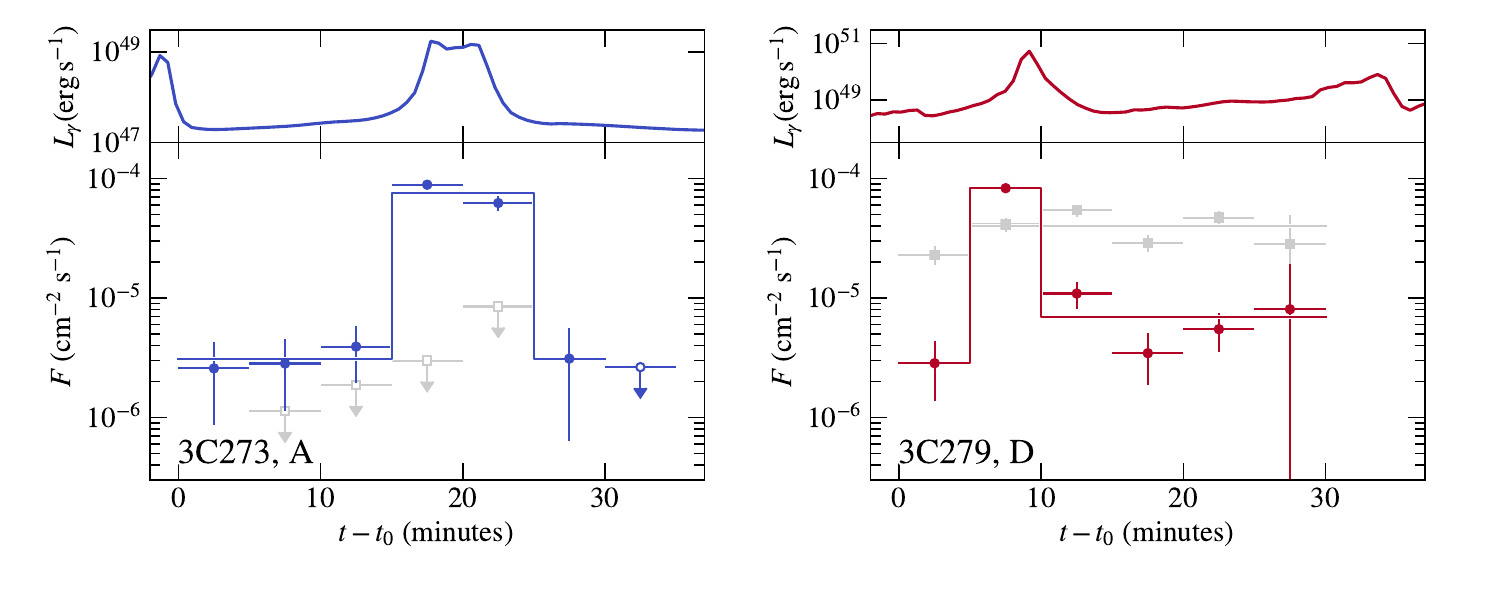}
    \caption{Theoretical and reconstructed light curves to search for variability on timescales of minutes. \textit{Left panels:} theoretical (top) and reconstructed light curves (bottom) for Model A and 3C~273. The theoretical light curve is shifted such that one short peak coincides with a GTI. 
    \textit{Right panels:} same as the upper panels but for Model D and 3C~279. Here,  $t_0=57,189.08$~MJD.}
    \label{fig:min-lc}
\end{figure*}{}

In order to test if even shorter variability time scales from reconnection  can be observed, we repeat the analysis for one good time interval (GTI), i.e., the time interval where the source is within the field of view of the satellite. 
Within one \emph{Fermi}-LAT orbit such a GTI has usually a duration of the order of $\sim30\,$minutes. 
We shift the theoretical light curve in time so that one of the pronounced sharp peaks falls inside a GTI and repeat our light-curve analysis, this time assuming a binning of 5\,minutes. 
We show the results for Models A and D in the left and right panels of Figure~\ref{fig:min-lc}, respectively. In both cases, the reconstructed light curves exhibit a luminous and fast flare, produced by one of the smaller and faster plasmoids in the layer. The peak flux of the flare above 100 MeV in both models is $\sim 10^{-4}\,\mathrm{s}^{-1}\,\mathrm{cm}^{-2}$. Such high fluxes have never been reported for 3C~273, while they are close to the highest flux ever reported for 3C~279 in 2015 June.
The actual observed \fermi-LAT light curve is shown with gray markers in the bottom right panel of Figure~\ref{fig:min-lc}. The reconstructed magnetic reconnection light curve reaches a flux about a factor of 2 higher than the observed flux, but only for one single 5~minute time bin. The minimum variability time obtained from the artificial light curve is $\min(t_\mathrm{var}) = (2.7\pm0.3)\,$minutes, which is about a factor of 3 shorter than the observed one, $\min(t_\mathrm{var}) = (8.2\pm2.6)\,$minutes. By tweaking the parameters of the magnetic reconnection simulation, such as the viewing angle of the layer, one can reduce the flux to the observed values and prolong the duration, thus having better agreement to the observations. 

One might worry that the radiation pressure of the plasmoid causing the fast variability is much higher than the ambient pressure in the jet. 
In the considered model $z_\mathrm{diss} \approx R_\mathrm{BLR}$ and the ambient jet pressure should be comparable to the pressure of the BLR~\citep[e.g.,][]{1984RvMP...56..255B}.  
For our model parameters of 3C~279, this is equal to $P_{\rm ext} \sim L_\mathrm{ext} / c z_{\rm diss}^2 \sim 0.2~\mathrm{dyne}\,\mathrm{cm}^{-2}$.
On the other hand, the observed luminosity for the plasmoid causing the fast variability in Model D is $L_\gamma \sim10^{50}\,\mathrm{erg}\,\mathrm{s}^{-1}$ averaged over the 5~minute time bin with the highest flux (see Figure~\ref{fig:min-lc}).
Omitting redshift factors, the radius of the plasmoid is limited by causality arguments to $r < c\Delta t$, which yields a lower limit on the pressure of plasmoid $P_p > L_\gamma / (\Delta t^2 c^3) \sim 5\times10^{13}\,\mathrm{dyne}\,\mathrm{cm}^{-2}$ for $\Delta t = 5\,$minutes.
The pressure in the plasmoid
co-moving frame is $P_p' = \delta_p^{-6} P_p$, where four powers come from the transformation of the luminosity and two powers from the transformation of $\Delta t$.
In order to have comparable pressures, one therefore has to require that $\delta_p \gtrsim (P_p / P_\mathrm{ext})^{1/6} \sim 250$.
This is satisfied for a plasmoid moving at perfect orientation for which $\delta_p \approx 4 \Gamma_j \Gamma_{\rm co} = 288 (\Gamma_j / 24) (\Gamma_{\rm co} / 3) \sim 288$ for our chosen model values of $\Gamma_j = 24$ and $\sigma = 10$.

\emph{Given the above results, we conclude conclude that magnetic reconnection events are a viable explanation for the observed short time variability reported in FSRQs.}

\begin{figure*}
    \centering
    \includegraphics[width=.99\linewidth]{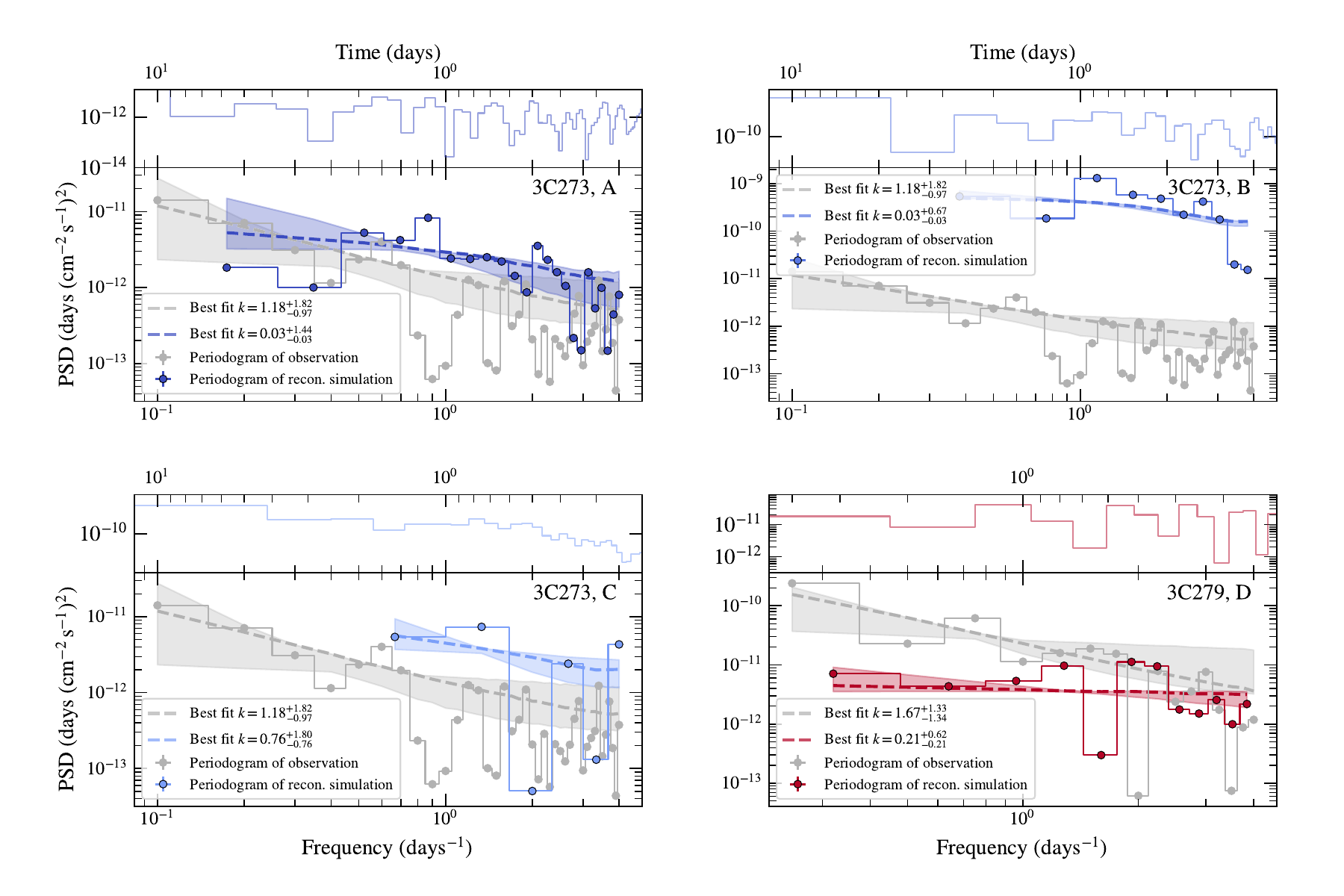}
    \caption{Periodograms for the theoretical, reconstructed, and observed light curves for all models. 
    Lower panels show the periodograms for the observed and reconstructed \fermi light curves.
    Upper panels show the periodograms for the theoretical light curves up to the maximum frequency of the simulated one.
    The best-fit values for $k_\mathrm{obs}$ and $k_\mathrm{art}$ are provided in the legend. 
    \label{fig:psd}}
\end{figure*}{}

\begin{deluxetable}{cc  cccc}
% \colnumbers
\centering
\tablecaption{Power-law slopes of PSDs for the theoretical, artificial, and observed \fermi-LAT light curves of 3C~273 and 3C~279.
\label{tab:results}
}
\tablewidth{0pt}
\tablehead{
\colhead{Model} & 
\colhead{Source} & 
\colhead{$k_\mathrm{theo}$\tablenotemark{\dag}} &  
\colhead{$k_\mathrm{theo}^{\mathrm{lim}\,\nu}$\tablenotemark{\ddag}} & 
\colhead{$k_\mathrm{art}$} &
\colhead{$k_\mathrm{obs}$}
}
\startdata 
A & 3C~273 & $1.306$ & $0.297$ & $< 1.47$ & $1.18^{+1.82}_{-0.97}$ \\
B & 3C~273 &  $2.075$ & $0.989$ & $<0.70$ & $1.18^{+1.82}_{-0.97}$\\
C & 3C~273 & $2.657$ & $1.126$ &$<2.56$ & $1.18^{+1.82}_{-0.97}$\\
D & 3C~279  & $1.451$ & $0.297$ & $<0.83$ & $1.67^{+1.33}_{-1.34}$ \\
\enddata
\tablenotetext{\dag}{These values are calculated from a simple log-log linear regression of the periodogram of the theoretical light curves shown in Figure~\ref{fig:theo-lc}.}
\tablenotetext{\ddag}{These values are also derived from a log-log linear regression of the periodogram of the theoretical light curves, however, with the frequency range limited to the ones of the simulated \fermi light curves shown in Figure~\ref{fig:lcs}.}
\end{deluxetable}

\subsection{Periodograms}
We also compute the periodograms of the 3 hr-binned observed and reconstructed light curves, which we fit with power spectral densities following power laws, $\mathrm{PSD}(\nu)\propto \nu^{-k}$, where the frequency $\nu$ is given by the inverse of time. 
We follow the methodology of  \citetalias{Meyer2019}, which is based on the work of \citet{2013MNRAS.433..907E} and \citet{2014MNRAS.445..437M} to fit the periodograms of the reconstructed simulated and observed light curves. 
In order to do so, we produce a sample of 
% artificial light curves  
fake light curves
(not to be confused with our artificial light curves obtained from the magnetic reconnection simulation)
for a grid of $k$ values, following the same flux distribution as the observed and  artificial light curves \citep{2013MNRAS.433..907E}. 
Furthermore, we apply the same time gaps to the fake 
% artificial
light curves as in the \fermi light curves (where $\mathrm{TS} < 9$).
In order to achieve an even sampling, the gaps are filled with values from a linear interpolation.
To avoid red-noise leakage, we simulate 100 light curves with a duration 100 times larger than the duration of the observed and reconstructed light curves, so that we end up with $10^4$ light curves for each input $k$ value with a length equal to the observed and 
artificial light curves.
As in \citetalias{Meyer2019}, we do not apply a window function.

The results for 3C~273 and 3C~279 are shown in Figure~\ref{fig:psd} for each magnetic reconnection model.
The best-fit average periodograms for the  artificial (colored dashed lines) and observed light curves (grey dashed lines) are shown with contours marking the $2\,\sigma$ uncertainties.
The corresponding $k_\mathrm{art}$ and $k_\mathrm{obs}$ values are reported in Table~\ref{tab:results}.
The best-fit values for all Models are compatible with $k_\mathrm{art} = 0$ and only 2\,$\sigma$ upper bounds can be derived. 
This is due to the short time duration of the magnetic reconnection event, which prevents us from  deriving the periodograms at lower frequencies, and the 3 hr binning applied to the light curves, which prevents the calculation of the periodogram at higher frequencies.
These small values of $k_\mathrm{art}$ are to be expected from the theoretical light curves.  When the latter are limited to the frequency range of the simulated \fermi observation, a simple linear regression in log-log space reveals periodograms with $k_\mathrm{theo}^{\mathrm{lim}\,\nu} \lesssim 1$, see Table~\ref{tab:results}. 
Meanwhile, the PSD of the full theoretical light curve has usually steeper slopes, ranging from  $k_{\rm theo} \sim 1.3$ to $k_{\rm theo} \sim 2.7$ (again using a simple log-log regression).  
If \fermi observations were able to also probe shorter time scales, the fitted periodograms should approach the one of the full theoretical light curves.
% , which have steeper slopes (i.e., higher values of $k_\mathrm{theo}$).
The best-fit $k_{\mathrm{art}}$ values of the reconstructed \fermi observations are smaller than the ones of the observed light curves, $k_{\mathrm{obs}}$. 
Given that the exact values for the slope depend on the specific time windows chosen here, and because of the large uncertainties on $k_{\rm obs}$, we cannot draw any firm conclusions by the comparison of the PSDs at the moment.

\subsection{BL Lac objects}

\begin{figure*}
    \centering
    \includegraphics[width=.99\linewidth]{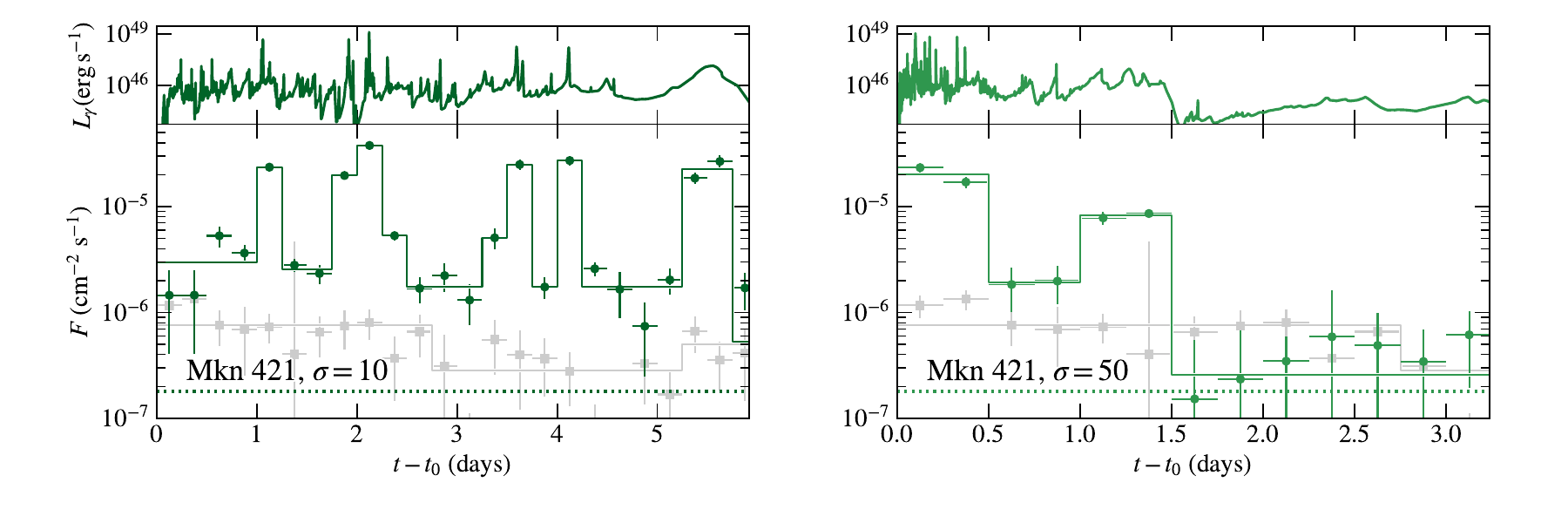}
    \caption{Same as Figure~\ref{fig:lcs} but for the case of the BL Lac Mkn~421 using the simulations of \citetalias{Christie2019} for parameter values tuned to BL Lac type sources with $\sigma=10$ (left) and $\sigma=50$ (right). 
    \label{fig:bllac}}
\end{figure*}

Although we have investigated the detectability of $\gamma$-ray flares from reconnection in bright FSRQs, our calculations can be extended to BL~Lac sources. Our reconnection model predicts broadband blazar spectra similar to those of BL Lac sources for higher jet magnetizations (i.e., $\sigma\gtrsim 10$) and for very weak or absent external radiation fields \citepalias{Christie2019}. We adopted the two vanilla BL Lac models of \citetalias{Christie2019} for $\sigma=10$ and 50, and created artificial LAT light curves of Mrk~421 for the time window MJD 56,124-56,131 as described in Section~\ref{sec:fakelc} (this time window coincides with the brightest $\gamma$-ray flare observed from this source with the LAT). 
The observed and artificial light curves are shown in Figure~\ref{fig:bllac}.
As can be seen, both models lead to source detections for a chosen 6~hr  binning, but produce higher fluxes (by almost one order of magnitude) and more structured light curves than observed.
Especially for the $\sigma=50$ case, an even finer binning could be chosen as well as the luminosity reaches levels close to $10^{49}\,\mathrm{erg}\,\mathrm{s}^{-1}$.
This would reveal variability on even shorter time scales.  
To match the variability properties and average flux level of the observed light curves of Mrk 421 in the LAT band, a lower $\Gamma_j$ (by a factor of $\sim 2$) or some misalignment between the observer and the reconnection layer would be required.

\section{Alternative models for fast $\gamma$-ray variability}\label{sec:models}

While emission from plasmoids in magnetic reconnection is a plausible explanation for the fast $\gamma$-ray variability detected in blazars by \fermi-LAT at gigaelectronvolt energies (\citealt{Ackermann2016}, \citetalias{Meyer2019}, \citealt{2020NatCo..11.4176S}) and by Cerenkov telescopes at very  high energies \citep[e.g.,][]{Aharonian2007, Albert2007}, a number of alternative models have been put forward to explain variability on minute-long timescales~\citep[for a detailed discussion on plausible models, see, e.g.][]{Aharonian2007,Aharonian2017}. 
Here, we discuss some of these models and potential ways they may be discriminated from reconnection scenarios.

For instance, \cite{Barkov2012} proposed that fast $\gamma$-ray variability can be produced by the interaction of a red giant (RG) star that crosses the blazar jet relatively close to the central black hole. Due to dynamic interactions with the jet, the RG can lose a significant fraction of its atmosphere. This is likely fragmented to smaller blobs (with co-moving radii $\mathcal{O}(10^{15})$~cm) that can be accelerated up to the jet Lorentz factor. 
Provided that the acceleration and cooling timescales of particles are much shorter than the typical dynamical timescale of the blob, $\gamma$-ray flares with observed durations of several minutes are expected. The profile of individual flares is mostly dictated by the dynamics of the jet-RG interaction. Similarly, in the magnetic reconnection scenario, the profile of each flare powered by a single plasmoid is mostly determined by its dynamics and motion in the layer. A more detailed comparison with the jet-RG scenario would require knowledge of the statistical properties of the fragmented stellar envelope. 
\cite{2014ApJ...780...87M} proposed that short timescale variability is produced by random changes in the density of relativistic electrons in a turbulent plasma crossing at relativistic speeds a standing conical shock in the parsec-scale blazar jet. In this model, the variations in flux are caused by continuous noise processes (described by PSD$(\nu) \propto \nu^{-k}$, $k=1.5-3$) rather than by a superposition of singular explosive events of energy injection at the base of the jet. The $\gamma$-ray flux (from Compton scattering of external photon fields) changes erratically on short timescales that are related to the Doppler factor of turbulent plasma cells containing the highest energy particles, randomly pointing toward the observer. The predicted X-ray flux (from synchrotron-self-Compton processes) usually shows smoother variations, as it is produced from lower energy particles occupying a larger volume. This difference between X-ray and $\gamma$-ray light curves is not expected in the magnetic reconnection scenario presented here, since the main cause of fast variability is mostly related to the relativistic motion of plasmoids in the jet. Another source of fast and bright $\gamma$-ray flares can be the magnetospheric gap close to the black hole event horizon \citep[e.g.,][]{Neronov2007, Levinson2000}. For low accretion rates, the injection of charges to the black hole magnetosphere is not sufficient for a full screening of the electric field that is induced by the rotation of the black hole. The acceleration of charged particles in unscreened electric fields (the so-called gaps) can be very efficient, leading to very high-energy emission that is variable on timescales $\sim h/c$; here, $h$ is the gap size and $h\le r_g \simeq 3\times10^{13}\left(M_{BH}/10^8 M_{\odot}\right)$~cm. These models have been applied mostly to radio galaxies \citep[e.g.,][]{Levinson2011, 2016A&A...593A...8P,2020ApJ...895...99K}, but are likely not relevant for bright quasars where no vacuum gaps are expected due to higher accretion rates. Other models for fast $\gamma$-ray variability  rely on very short cooling timescales for the $\gamma$-ray emitting electrons. High radiative efficiency can be achieved if the radiating particles are ultra-relativistic pairs produced by proton-photon interactions, radiating via synchrotron in gigaelectronvolt $\gamma$-rays (e.g., \citealt{Ackermann2016}, \citealt{Shukla2018}, \citetalias{Meyer2019}). These models, however, have not matured enough, as to make predictions for the multiwavelength emission and variability properties of $\gamma$-ray emission (e.g., PSDs).

\section{Summary and Discussion} \label{sec:discussion} 
% \MP{[Because the discussion will become too long, I think it makes sense to divide it in thematic subsections. I gave it a try, but feel free to modify it]}

We have simulated artificial $\gamma$-ray light curves from a chain of plasmoids formed during a magnetic reconnection event in the blazar jet.
These light curves have been used as input for the simulation of artificial light curves that would be observed with the \textit{Fermi}~LAT.
As test beds, we have used observations of two FSRQs, 3C~273 and 3C~279, that encompass reported $\gamma$-ray flares. 
From a qualitative comparison of these reconstructed \textit{Fermi}-LAT light curves to observed $\gamma$-ray flares, we can draw the following conclusions.
When a 3~hr binning of the LAT light curves is adopted to guarantee source detections of these bright FSRQs in each time bin, much of the rapid variability of the theoretical plasmoid light curves is washed out (see Figures~\ref{fig:theo-lc} and \ref{fig:lcs}).
However, the average flux levels, the minimum variability time scales, and the overall structure of the light curves are comparable. 
In particular, the plasmoid simulations with a slight misalignment between the reconnection layer give rise to broader features in the light curve compared to the fully aligned case. 
This is in better agreement with the observed flare of 3C~273 around MJD~55,094.  

\subsection{Appearance of artificial LAT light curves and their PSDs}
The model-predicted variability is highly dependent upon the orientation of the reconnection layer with respect to the blazar jet axis and to the observer (see Models B and C in Figure~\ref{fig:theo-lc}). For optimal orientations, the theoretical light curves exhibit fast and powerful flares powered mostly by small and fast plasmoids. 
These sharp, spike-like outbursts of emission appear as excesses atop a more slowly evolving envelope produced by the cumulative emission of medium-sized plasmoids.
These outbursts can easily explain minute-scale variability in FSRQs and are a clear-cut prediction of our reconnection model (see Figure~\ref{fig:min-lc}).
However, these outbursts are easily washed out when a coarse binning of the light curves is chosen. Thus, a systematic search of minute-scale flares on sub-orbital time scales could serve as a test for magnetic reconnection events in blazar jets.
It has long been predicted that magnetic reconnection can give rise to fast flares on top of a slowly varying flux \citep{Giannios2013}, which could explain evidence for fast minute-scale variability \citepalias[e.g.,][]{Christie2019}.
To our knowledge, we have been able to show for the first time that such fast variability can indeed be observed with the \fermi~LAT.

A recent analysis of \fermi-LAT data of 3C~279 from 2018 led to the identification of fast flares (on minute-long time scales) on top of a slower varying envelope \citep{2020NatCo..11.4176S}. These observations were put forward as evidence of magnetic reconnection occurring in blazar jets, although they were discussed only on a qualitative level. The findings presented in this work, which are derived using state-of-the-art simulated $\gamma$-ray light curves and \fermi-LAT analysis tools, put the magnetic reconnection scenario for fast $\gamma$-ray variability in blazars on a more firm basis.

\begin{figure}
    \centering
    \includegraphics[width=.99\linewidth]{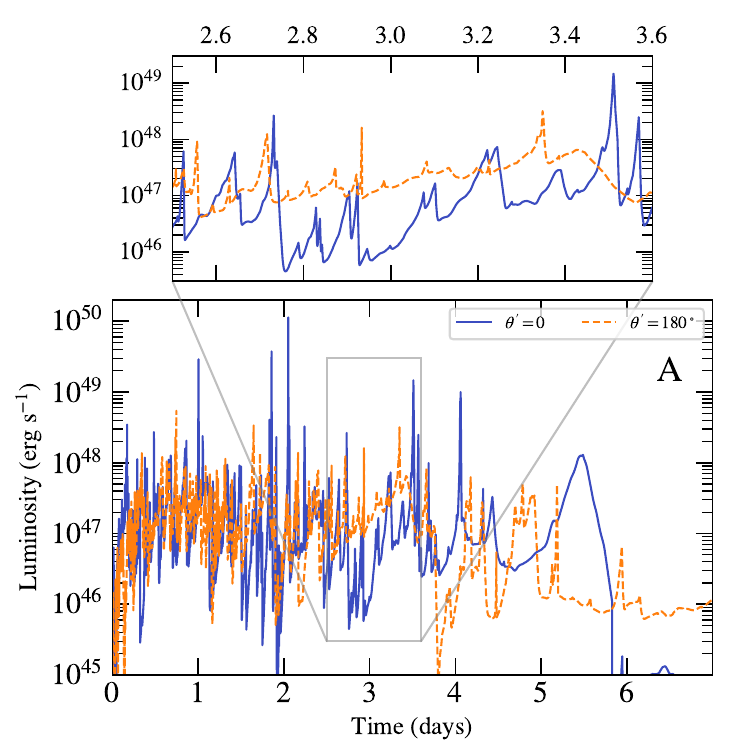}
    \caption{
    Theoretical $\gamma$-ray light curves from magnetic reconnection in Model A with different layer orientations. The blue line is the same as Model A in Figure~\ref{fig:theo-lc}, for which $\theta' = 0^\circ$.
    The orange dashed line shows the same simulation but with a layer orientation of $\theta' = 180^\circ$. 
    The inset shows a zoom-in of the time where the monster plasmoid dominates the emission for the $\theta' = 180^\circ$ case. Fast flares on top of the slowly rising emission from the monster plasmoid are produced by smaller and fast plasmoids in the layer.
    }
    \label{fig:comp-0-pi}
\end{figure}

In all models discussed so far, the $\gamma$-ray light curves from reconnection events end
% seize
with a large-amplitude flare stretched out in time (see Figures~\ref{fig:theo-lc} and \ref{fig:lcs}). This flare can be attributed to a single slow plasmoid that has undergone significant growth to a sizable fraction of the reconnection region \citep{Uzdensky2010}. Only one or two such ``monster'' plasmoids are expected to reside in the layer at any given time~\citepalias[see, e.g., red colored curves in Fig.~1 of][]{Christie2019}. Despite their low Doppler factors,\footnote{The largest plasmoids in the layer are moving with mildly relativistic or nonrelativistic speeds in the layer's frame \citep{Sironi2016}, i.e. the jet's co-moving frame.}, monster plasmoids produce flares with similar bolometric luminosities as those powered by smaller and faster plasmoids due to the larger number of contained radiating particles \citep{Petropoulou2016}; for instance, compare the photon fluxes of the peaks in the reconstructed light curve of Model A in Figure~\ref{fig:lcs}. The emergence of a long-duration and luminous flare toward the end of a reconnection event is not, however, a universal feature of magnetic reconnection. Such flares can be produced at earlier times, depending on the relative orientation of the observer and reconnection layer. For the orientations considered here, an observer receives Doppler-boosted emission from plasmoids formed in the right side of the reconnection layer, where only one monster plasmoid is formed and exits the layer at late times. If instead $\theta_{\rm obs}=0^{\rm o}$ and $\theta^\prime=180^{\rm o}$, the observer would receive Doppler-boosted radiation from plasmoids in the left side of the layer, where a monster plasmoid is formed well before the end of the reconnection event \citepalias[see the red curve for $x<0$ in the middle panel of Figure~1 in][]{Christie2019}. For illustration purposes, the theoretical light curve of Model A for $\theta^\prime=180^{\rm o}$ is shown in Figure~\ref{fig:comp-0-pi}. For an on-axis observer, monster plasmoids would lead to flares of similar peak luminosity for almost all angles $\theta' \in [0^{\rm o},180^{\rm o}]$ \citep{Petropoulou2016}. Several luminous flares with long durations may be expected from a single reconnection event, since monster plasmoids are ejected from the current sheet every $\sim 3 L/c$ (as measured in the rest frame) as shown in the top panel of Figure~9 in  \citet[][]{Sironi2016}.

We have also investigated the PSDs of the theoretical light curves and the simulated \textit{Fermi}-LAT light curves. 
In the limited frequency range of these light curves, the PSDs can be well described by simple power laws with spectral indices $k < 1$.
Within uncertainties, these PSDs agree with the observed ones.
In order to draw firm conclusions in the future, magnetic reconnection simulations over longer time scales would be highly desirable. 
Additionally, a PSD analysis of unbinned \textit{Fermi} data could help to probe frequencies below the chosen time binning \citep[e.g.,][]{2019ApJ...885...92K}. All FSRQ models  discussed so far were based on simulations of reconnection in plasmas with fixed magnetization $\sigma=10$. The variability properties of the theoretical light curves depend on $\sigma$  though \citepalias{Christie2019}  (see also the theoretical light curves of Mrk~421 in Fig.~\ref{fig:bllac}). More specifically, an increase of $\sigma$ alone results in the production of more small and fast plasmoids within the layer, yielding a brighter and more variable light curve \citep[see also][]{Sironi2016}. As a result, the theoretical PSD slopes should also be $\sigma$ dependent. As an indicative  example, we computed the light curves of Model A using PIC simulation results for $\sigma=3$  and 50 and found PSDs with much softer and harder slopes $k_{\rm theo}$, respectively. 
For $\sigma = 3$ we find $k_{\rm theo} = 2.97$ whereas $\sigma = 50$ results in $k_{\rm theo} = 0.19$. 
Thus, increasing $\sigma$ shifts the PSDs from red noise to pink noise and eventually close to white noise. 
For a well-sampled light curve, the PSDs could in principle be used to infer the magnetization of the unreconnected plasma in the jet. 

\subsection{Origin of external photon fields}
With our chosen simulation parameters, it is necessary to assume an external photon field luminosity of $L_\mathrm{ext} =10^{46}\,\mathrm{erg}\,\mathrm{s}^{-1}$ in order to 
reproduce the $\gamma$-ray flare flux of 3C~279. Still, the average flux of the reconstructed light curve over a time window of 3 days is $0.31 \pm 0.01$ times lower than the observed one (see Model D in Figure~\ref{fig:lcs}).
If the external photon field is provided by the BLR, we need to set the H$\beta$ luminosity in the model of \citet{Finke2016} to $3.3\times10^{44}\,\mathrm{erg}\,\mathrm{s}^{-1}$. 
This is about 20 times brighter than the H$\beta$ luminosity observed from 3C~279  \citep{2006ApJ...637..669L}. 
This discrepancy might be explained by a brightening of the BLR; flare-like activity of emission lines from the BLR has been found in coincidence with enhanced nonthermal emission from the jet of 3C 454.3 and CTA~102 \citep{2013ApJ...763L..36L, 2020ApJ...891...68C}.
The flux from the Mg~II line was found to vary by a factor of a few, whereas the ratio between maximum and minimum flux for the Fe~II line in CTA~102 reached a value of 34.
Additionally, \citet{2020ApJ...891...68C} interpreted their results as evidence for an additional more extended BLR component, which would provide an additional source of external photons. Being more diluted ($u_{\rm ext} \propto R^{-2}_{\rm ext}$) though, this extra BLR component would not be an important source of seed photons for inverse Compton scattering (ICS). However, if the main BLR component was more extended than what the simple scaling relation with the disk luminosity predicts, then the dissipation site could be pushed to larger distances, thus relaxing the constraint on the jet bulk Lorentz factor for same layer length $L$ forming in a conical jet ($z_{\rm diss} \approx \Gamma_j L \leqslant R_{\rm ext}$). 
Since the total inverse Compton power scales as $\Gamma_j^{2+p} L^2 R_{\rm ext}^{-2}$ (see equation \ref{eq:Ptot_gamma_j}), a higher  Lorentz factor would increase the inverse Compton emission considerably. 
We have also checked that smaller $L_\mathrm{ext}$ values, which are compatible with the values reported by \citet{2006ApJ...637..669L}, result in very low $\gamma$-ray fluxes that cannot reproduce the flare flux of 3C~279 in 3~hr time bins (the power of the emission produced by external Compton scattering scales linearly with $L_\mathrm{ext}$, see the Appendix).

\subsection{Model assumptions and caveats}
While the assumptions of the model used for the construction of artificial LAT light curves are described in detail in \citetalias{Christie2019}, we report here those that are most relevant to this work. We also discuss possible caveats and their impact on our results.

Our radiative model is benchmarked with 2D PIC simulations that dictate the bulk motion of plasmoids in the layer and the properties of accelerated particles in the absence of a guide field\footnote{The magnetic field component normal to the plane of reconnection.}. 
Reconnection in the presence of strong guide fields (with comparable to or larger strengths than that of the alternating magnetic field) is known to be less efficient in producing relativistic particles,  while resulting in plasmoids with more magnetically dominated interiors \citep[e.g.,][]{Sironi2015}. In this case, the emission from plasmoids would be less luminous and more synchrotron dominated than the one found in our radiative model for the same parameters.

For the purposes of the radiative transfer calculations, plasmoids were assumed to be spherical structures. Three-dimensional (3D) PIC simulations of reconnection have shown that the evolution of a 3D antiparallel current sheet results in elongated flux tubes that grow and coalesce over time, similar to the 2D physics \citep[e.g.,][]{2008ApJ...677..530Z, 2014PhRvL.113o5005G,sironi2014S,2017ApJ...843L..27W}. Still, the timescales and spatial domains covered in 3D PIC simulations of reconnection are not long and large enough yet to allow a detailed study of the evolution of these structures as in 2D \citep[e.g.,][]{Sironi2016, 2019ApJ...880...37P}. Despite the presence of the drift-kink instability, which can corrugate the current sheet \citep[e.g.,][]{2007ApJ...670..702Z, 2008ApJ...677..530Z, 2014PhRvL.113o5005G}, particle acceleration is not suppressed in 3D; in fact, the nonthermal particle spectra resemble those found in 2D PIC simulations \citep{sironi2014S,2017ApJ...843L..27W}. As in 2D, the presence of a strong guide field in 3D reconnection can suppress particle acceleration  leading to steeper particle spectra \citep{2017ApJ...843L..27W}. These results show that 2D PIC studies are still pertinent for 3D reconnection. 

Our reconnection model assumes a slab geometry for the current sheet forming in the blazar jet. Moreover, the properties of the reconnection layer (e.g., length and distance from the black hole) are considered  free parameters. In a more realistic model, the current sheet properties should be dictated by the large-scale dynamics of the jet and the initial conditions at the jet base. For instance, a jet with alternating toroidal field polarity along its propagation axis (striped jet) could form when the open magnetic field lines threading the black hole horizon reverse their polarity \citep[e.g.,][]{2015MNRAS.446L..61P, 2019MNRAS.484.1378G, 2019MNRAS.490.4811C}. In such a scenario, the assumption of slab geometry is appropriate. However, current sheets with helicoidal shapes are expected if magnetic reconnection is driven by kink instabilities in jets with helical magnetic fields \citep[e.g.,][]{2017ApJ...835..125Z, 2021MNRAS.501.2836B}. In the case of turbulence-driven reconnection, current sheets are short-lived, acting as locations of fast energy gain for particles that are then stochastically accelerated by the magnetic turbulence \citep[][and references therein]{2019ApJ...886..122C}. In conclusion, the radiative model used in this work is mostly relevant to reconnection with weak guide fields and long-lived current sheets.

As mentioned in Section~\ref{sec:theory}, particles are assumed to be accelerated in the current sheet into a power-law momentum distribution with index $p$ that depends on $\sigma$ before being injected into plasmoids. The power-law index used in the radiative model has been measured from 2D PIC simulations using the particle spectrum extracted from the whole reconnection region and not from individual plasmoids. Nonetheless, this is a good assumption especially for medium and large plasmoids whose particle spectrum has been shown to have similar slope as that of the total spectrum \citep{Sironi2016, Petropoulou2018}. Still, our radiation model does not take into account other secondary effects found in PIC simulations. For instance, softening of the total power-law particle spectrum toward its asymptotic value $p$ has been reported by \citet[][see Figures~3 and~6]{Petropoulou2018} at early times (with respect to the onset of the reconnection event). A time-dependent power-law index could contribute to the spectral variability of the synchrotron and SSC components. Moreover, particles can gain energy during plasmoid mergers \citep[e.g.,][]{sironi2014S, 2015ApJ...815..101N} that occur after their injection into plasmoids.  In our radiative model, a merger could be described by an instantaneous episode of injection of particles with a bias toward high energies. Although the number  of  particles  accelerated  during  a  merger  may  be small compared  to  those  accelerated  at  X-points in the current sheet, a merger may lead to flares with spectral hardening. Such effects would be more prominent though in energy bands where the highest energy particles radiate. Finally, the role of slower processes of particle energization within plasmoids (i.e., adiabatic compression) has been only recently recognized  in relativistic reconnection \citep{Petropoulou2018,Hakobyan2020}, and merits a separate study.
 
In our radiative model, plasmoids are assumed to be isolated objects and the particles contained within are shielded from the radiation of other plasmoids. This additional radiation could act as an external force on plasmoids (Compton drag) opposing  the magnetic tension force that is mostly responsible for the bulk acceleration of plasmoids in the current sheet \citep{Beloborodov2017}.
% , but for the case of an ambient intense photon field. 
However, it has been recently demonstrated with radiative 2D PIC simulations of  reconnection in pair plasmas that the bulk deceleration of small plasmoids is not severe \citep[see Figure~5 in][]{sironi2020}. In fact, the removal of a plasmoid's internal energy due to Compton cooling of particles is expected to happen faster than the removal of its bulk energy due to Compton dragging. Hence, the Doppler beaming in our radiative calculations, which is important for the predicted fast variability, would not be much affected by the Compton drag. Photons emitted by the large and slow plasmoids in the layer are an additional source of soft photons that can be Compton upscattered by particles in small fast-moving neighboring plasmoids. This inter-plasmoid Compton scattering process, which has been studied by \citet{christie2020_IPCS},  can naturally occur throughout the reconnection layer and enhance the Compton flux by a factor of a few. This process, however, is expected to be important for blazars without strong ambient radiation fields.

Once plasmoids merge or advect from the layer, PIC simulations can no longer track their properties and growth. As such, we assume that at these points in time, particle injection ceases abruptly, leaving the remaining particles within a given plasmoid to cool within the ambient external radiation fields. Although this has a direct consequence on the decay phase of the resulting light curve, it should be noted that in extreme radiation fields, as those encountered in bright FSRQs, particles within fast-moving plasmoids are typically cooled well before the merger/advection time (see also the next paragraph). As a result, the decaying part of the flare depends mostly on the way fresh particles are injected post-merger or post-advection.  \cite{Petropoulou2016} computed numerically the radiation from  a single plasmoid considering an exponentially decaying particle injection rate post-merger or post-advection (see equation 41 therein), and a power-law decay of the magnetic field 
Unless particle injection lasts for several light-crossing times of the plasmoid (measured before merger/advection), the decay time of the flare is not going to be much longer than its rising time (see, e.g., Figure 11 of \citealt{Petropoulou2016}).

Plasmoids forming close to $R_{\rm ext}$ could, within their lifetime, travel beyond $R_{\rm ext}$, thereby reducing the photon energy density as seen in their rest frame and effectively washing out the luminous $\gamma$-ray emission. Fast plasmoids, which are responsible for producing the shortest flares in the light curves, typically have lifetimes (as measured in the jet frame) that are a fraction of $L/c$ \citep[see, e.g., Figures 4 and 8 in][]{Sironi2016}. Meanwhile, the peak of plasmoid emission typically occurs at a fraction of their lifetime \citepalias{Christie2019}. Thus, the observable time window of a plasmoid is effectively shorter than its lifetime by a factor of  $f<1$. For instance, the distance traveled by fast plasmoids in our simulations (in the galaxy frame) will be $d \approx \Gamma_p f (L/c) c \simeq 3.6\times10^{17}~{\rm cm}\, (f/0.1) (\Gamma_p/72)(L/5\times10^{16}{\rm cm})$.

\section{Conclusion}\label{sec:conclusion}
We have shown that artificial \textit{Fermi}-LAT light curves generated from magnetic reconnection simulations are in general able to reproduce the characteristics of observed $\gamma$-ray flares of FSRQs.
Such characteristics include the average flux level and minimum variability time scale. 
Also the power spectral densities of the  artificial and observed light curves are compatible. 
We have further been able to show that it is  possible to explain the observed fast minute-scale variability in FSRQs with fast plasmoids moving close to the line of sight. 
However, the times of the fast spikes in the magnetic reconnection simulations need to coincide with the times when the source was in the field of view of the LAT.

These promising results motivate future work. From a theoretical perspective, it is desirable to compute the emission produced from multiple reconnection layers formed at different distances within the jet in an attempt to simulate the average $\gamma$-ray blazar emission (i.e., the QB). Extension of our radiative transfer calculations to times after the plasmoid exit from the layer, where they may undergo adiabatic expansion,  would be necessary for computing a  likely delayed radio signal following fast $\gamma$-ray flares. From an observational perspective, a systematic search for short outbursts in \textit{Fermi}-LAT data could provide further evidence for the presence of fast-moving plasmoids in blazar jets. 

\acknowledgments
The authors would like to thank the anonymous referees for their constructive comments.
The authors would also like to thank Justin Finke, Sara Buson, Matthew Kerr, Philippe Bruel, and Roger Blandford for helpful discussions and comments on the manuscript. 
M. M. acknowledges support from the Alexander von Humboldt  Foundation and the  European  Union's  Horizon 2020 research and innovation program under the Marie Sk{\l}odowska-Curie grant agreement GammaRayCascades No~843800.
MP acknowledges support from the Lyman Jr. Spitzer Postdoctoral Fellowship. M. P. and I. C. are supported by the Fermi Guest Investigation grant 80NSSC18K1745. 

The \textit{Fermi} LAT Collaboration acknowledges generous ongoing support
from a number of agencies and institutes that have supported both the
development and the operation of the LAT as well as scientific data analysis.
These include the National Aeronautics and Space Administration and the
Department of Energy in the United States, the Commissariat \`a l'Energie Atomique
and the Centre National de la Recherche Scientifique / Institut National de Physique
Nucl\'eaire et de Physique des Particules in France, the Agenzia Spaziale Italiana
and the Istituto Nazionale di Fisica Nucleare in Italy, the Ministry of Education,
Culture, Sports, Science and Technology (MEXT), High Energy Accelerator Research
Organization (KEK) and Japan Aerospace Exploration Agency (JAXA) in Japan, and
the K.~A.~Wallenberg Foundation, the Swedish Research Council and the
Swedish National Space Board in Sweden.
 
Additional support for science analysis during the operations phase is gratefully
acknowledged from the Istituto Nazionale di Astrofisica in Italy and the Centre
National d'\'Etudes Spatiales in France. This work performed in part under DOE
Contract DE-AC02-76SF00515.

\bibliography{mainbib}

\appendix

\section{External Compton Power from a Single Plasmoid}
In this section we provide analytical expressions for the external Compton power emitted by  fast cooling electrons in plasmoids formed in the reconnection layer.

The single-particle ICS power (in the Thomson regime) is
\eqb
\label{eq:PEC}
P_{\rm EC}^\prime \approx \frac{4}{3} \, \sigma_T \, c {\gamma^\prime}^2  \, u_{\rm ext}^\prime,
\eqe
where a prime denotes any quantity measured in the co-moving frame of the plasmoid.  If the main source of seed photons for ICS is an external radiation field (external Compton; EC), like the BLR, then the photon energy density $u_{\rm ext}^\prime$ is written as \citep{Ghisellini1996}
\eqb
\label{eq:Uph}
u_{\rm ext}^\prime \approx \frac{17 \,  \Gamma_p^2 L_{\rm ext}}{48 \, \pi \, R_{\rm ext}^2 c},
\eqe
where $\Gamma_p$ is the plasmoid Lorentz factor
in the galaxy frame
\citep[see equation 6 in][]{Petropoulou2016}, $L_{\rm ext}$ is the bolometric luminosity of the external radiation field, and $R_{\rm ext}$ is its radial distance from the black hole (assuming a spherical shell). 

Substitution of Equation (\ref{eq:Uph}) into (\ref{eq:PEC})  yields 
\eqb
P_{\rm EC}^\prime \approx \frac{17 \, \sigma_T \, L_{\rm ext}\, \Gamma_p^2 {\gamma^\prime}^2}{ 36 \, \pi \, R_{\rm ext}^2}.
\eqe
As long as ICS on the external photons dominates particle cooling, which is a valid assumption for luminous FSRQs, the electron cooling Lorentz factor\footnote{This is defined as the Lorentz factor to which an electron cools in one light-crossing time of the plasmoid in its co-moving frame.} can be approximately written as
\eqb 
\label{eq:gcool}
\gamma^\prime_c \approx \frac{3 m_e c^2}{2 \sigma_T u^\prime_{\rm ext} w_\perp},
\eqe 
where $w_\perp$ is the plasmoid transverse diameter (i.e., perpendicular to the layer) and can be expressed as a fraction of the layer's half length $L$. Because of the plasmoid growth and plasmoid acceleration along the layer \citep{Sironi2016}, $\gamma^\prime_c$ also depends on time through the terms $w_\perp$ and $u^\prime_{\rm ext}$. For a semi-analytical derivation of the time-dependent expressions of $w_\perp$ and $\Gamma_p$ we refer the reader to \cite{Petropoulou2016}.

% \subsection{Fast-Cooling Regime}
% \label{sec:fast_cooling}
In the fast cooling regime, i.e., when $\gamma_c^\prime <\gamma^\prime_{\min}$,
the particle distribution $N_e(\gamma^\prime, t^\prime)$ within each plasmoid can be expressed as
\eqb
\label{eq:Ne_fast_cooling}
N_e (\gamma^\prime, t^\prime) = N_0(t^\prime)
\begin{cases}
{\gamma^\prime}^{-2}, \, \quad {\gamma^\prime}_{\rm c} < {\gamma^\prime} \le {\gamma^\prime}_{\rm min}\\ \\
{\gamma^\prime}^{-p-1} \, {\gamma^\prime}_{\rm min}^{p-1}, \quad {\gamma^\prime}_{\rm min } < {\gamma^\prime} \le {\gamma^\prime}_{\rm max}
\end{cases},
\eqe
where $\gamma^\prime_{\min}, \gamma^\prime_{\rm max}$ are the minimum and maximum Lorentz factors of the injected particles. Here, $p$ denotes the slope of the injected power-law distribution of particles, taken to be dependent upon the magnetization $\sigma$ as governed by PIC simulations \cite[e.g.,][]{sironi2014S}.

The energy-independent part of the particle distribution function, $N_0(t^\prime)$, can be determined using particle conservation within each plasmoid
\eqb
\label{eq:conserve}
\int {\rm d}{\gamma^\prime} \, N_e(\gamma^\prime, t^\prime) = \int_0^{t^\prime} {\rm d}\Tilde{t}^\prime \, \int_{{\gamma^\prime}_{\rm min}}^{{\gamma^\prime}_{\rm max}} {\rm d}{\gamma^\prime} \, Q_{\rm inj}(\gamma^\prime, \Tilde{t}^\prime) \equiv N_{\rm inj} (t^\prime).
\eqe
where the total number of  particles injected in the plasmoid volume $V(t^\prime)\propto w_\perp^3(t^\prime)$ by time $t^\prime$ is $N_{\rm inj} (t^\prime) \approx U_B^\prime \, n_{\rm PIC} \, V(t^\prime) / (2 \, \sigma \, m_p \, c^2)$ \citepalias{Christie2019}. Here, $n_{\rm PIC}$ is the area-averaged particle number density per plasmoid as determined from PIC simulations \citep[see panels (d)-(f) in Figure~5 of ][]{Sironi2016}, $U_B^\prime = B^{\prime 2} / 8 \pi$ is the magnetic energy density in the plasmoid co-moving frame, and $Q_{\rm inj}$ is the injection rate of accelerated particles. 
Using Equations (\ref{eq:Ne_fast_cooling})-(\ref{eq:conserve}), we find $N_0 (t^\prime) = N_{\rm inj} (t^\prime)/g(t^\prime)$ where 
\eqb
\label{eq:gt_fast_cooling}
g(t^\prime) = {\gamma^\prime}_c(t^\prime)^{-1} - {\gamma^\prime}_{\rm min}^{-1} + \frac{{\gamma^\prime}_{\rm min}^{p-1}}{p} ({\gamma^\prime}_{\rm min}^{-p} - {\gamma^\prime}_{\rm max}^{-p}) \approx \gamma_c^{\prime -1}\left(1-\frac{(p-1)\gamma_c^\prime }{p\gamma^\prime_{\min}} \right).
\eqe
and $\gamma^\prime_{\max}\gg \gamma^\prime_{\min}$ was assumed to obtain the right-hand side of the equation.

To compute the EC power of the plasmoid in a specific frequency range [$\nu_1, \nu_2$], we convolve $N_e$ with $P_{\rm EC}^\prime$, 
\eqb
P_{\rm EC, tot}^\prime &=& \int_{\gamma_1^\prime}^{\gamma_2^\prime} {\rm d}\gamma \, P_{\rm EC}^\prime \, N_e (\gamma^\prime, t^\prime)\\
& \approx & \frac{17 \, \sigma_T \, L_{\rm ext} \, \Gamma_p^2 \, N_{\rm inj}(t^\prime) \, {\gamma^\prime}_{\rm min}^{p - 1}}{36 \, \pi \, R_{\rm ext}^2 \, (2 - p) \, g(t^\prime)} ({\gamma_2^\prime}^{2-p} - {\gamma_1^\prime}^{2-p}),
\eqe
where the last expression was derived after assuming that $\gamma^\prime_{1,2} > \gamma^\prime_{\rm min}$. Here, $\gamma^\prime_{1,2} \approx  \sqrt{\nu_{1,2}/(\delta_p \Gamma_p \nu_{\rm BLR})}$ are the Lorentz factors of particles up-scattering external photons of typical frequency $\nu_{\rm ext}\approx 3k T_{\rm ext}/h$ to $\gamma$-ray frequencies $\nu_{1,2}$, and  $\delta_p$ is the plasmoid Doppler factor (see equation 8 in \citet{Petropoulou2016}).  Inserting the expression of $\gamma^\prime_{1,2}$ in the equation above, and using the relation for the observed EC power, $P_{\rm EC, tot} \approx \delta_p^4 \, P_{\rm EC, tot}^\prime$, we obtain
\eqb
\label{eq:Ptot}
P_{\rm EC, tot} \approx \frac{17 \, \sigma_T \, L_{\rm ext} \, \Gamma_p^{(2 + p)/2} \, N_{\rm inj}(t^\prime) \, \delta_p^{(6 + p)/2} \, {\gamma^\prime}_{\rm min}^{p - 1}}{36 \, \pi \, R_{\rm ext}^2 \, (2 - p) \, \nu_{\rm ext}^{(2-p)/2} \, g(t^\prime)} (\nu_2^{(2-p)/2} - \nu_1^{(2-p)/2}).
\eqe
For perfect orientation (i.e., $\theta_{\rm obs} = \theta^\prime = 0^\circ$), $\Gamma_p \approx 2 \Gamma_{\rm co} \Gamma_j$, and  $\delta_p \approx 4 \Gamma_{\rm co} \, \Gamma_j$ \citep{Petropoulou2016}, where $\Gamma_{\rm co}\lesssim \sqrt{\sigma}$ is the plasmoid Lorentz factor in the frame of the reconnection layer. In this case, equation (\ref{eq:Ptot}) simplifies to
\eqb
\label{eq:Ptot_gamma_j}
P_{\rm EC, tot} \approx \frac{17 \, \sigma_T \, 4^{(6+p)/2} \, L_{\rm ext} \, (\Gamma_j \Gamma_{\rm co})^{4 + p} \, U_B^\prime \, n_{\rm PIC} \, V(t^\prime) \, {\gamma^\prime}_{\rm min}^{p - 1}}{72 \, \pi \, \sigma \, m_p c^2 \, R_{\rm ext}^2 \, (2 - p) \, \nu_{\rm ext}^{(2-p)/2} \, g(t^\prime)} (\nu_2^{(2-p)/2} - \nu_1^{(2-p)/2}) \propto \Gamma_j^{2+p} L^2.
\eqe
where we used the relation $g(t')\propto \gamma^{\prime -1}_c \propto w_\perp u^\prime_{\rm ext}\propto L \Gamma_p^2$.
\end{document}